\documentclass{ws-ijmpa}
\usepackage{graphicx}

\newcommand\rf[1]{(\ref{eq:#1})}
\newcommand\lab[1]{\label{eq:#1}}
\newcommand\nonu{\nonumber}
\newcommand\br{\begin{eqnarray}}
\newcommand\er{\end{eqnarray}}
\newcommand\be{\begin{equation}}
\newcommand\ee{\end{equation}}

\newcommand\foot[1]{\footnotemark\footnotetext{#1}}
\newcommand\lb{\lbrack}
\newcommand\rb{\rbrack}
\newcommand\llangle{\left\langle}
\newcommand\rrangle{\right\rangle}

\newcommand\llb{\left\lbrack}
\newcommand\rrb{\right\rbrack}

\renewcommand\({\left(}
\renewcommand\){\right)}
\newcommand\bv{\bigm\vert}               

\newcommand\bc{\begin{center}}
\newcommand\ec{\end{center}}

\newcommand\partder[2]{\frac{{\partial {#1}}}{{\partial {#2}}}}

\renewcommand\a{\alpha}
\renewcommand\b{\beta}

\renewcommand\d{\delta}

\newcommand\vareps{\varepsilon}
\newcommand\g{\gamma}
\newcommand\G{\Gamma}

\newcommand\h{\frac{1}{2}}
\renewcommand\k{\kappa}
\renewcommand\l{\lambda}
\renewcommand\L{\Lambda}
\newcommand\m{\mu}
\newcommand\n{\nu}

\newcommand\vp{\varphi}

\newcommand\pa{\partial}

\newcommand\s{\sigma}
\renewcommand\S{\Sigma}
\renewcommand\t{\tau}
\renewcommand\th{\theta}

\newcommand\cA{{\mathcal A}}

\newcommand\cE{{\mathcal E}}
\newcommand\cF{{\mathcal F}}

\newcommand\cJ{{\mathcal J}}

\newcommand\cM{{\mathcal M}}

\newcommand\cV{{\mathcal V}}

\newcommand{\ct}[1]{\cite{#1}}
\newcommand{\bib}[1]{\bibitem{#1}}
\newcommand\PRL[3]{\textsl{Phys. Rev. Lett.} \textbf{#1}, #3 (#2)}
\newcommand\NPB[3]{\textsl{Nucl. Phys.} \textbf{B#1}, #3 (#2)}

\newcommand\PRD[3]{\textsl{Phys. Rev.} \textbf{D#1}, #3 (#2)}

\newcommand\PLB[3]{\textsl{Phys. Lett.} \textbf{#1B}, #3 (#2)}
\newcommand\CQG[3]{\textsl{Class. Quantum Grav.} \textbf{#1}, #3 (#2)}

\newcommand\AoP[3]{\textsl{Ann. of Phys.} \textbf{#1}, #3 (#2)}
\newcommand\RMP[3]{\textsl{Rev. Mod. Phys.} \textbf{#1}, #3 (#2)}

\newcommand\IJMPA[3]{\textsl{Int. J. Mod. Phys.} \textbf{A#1}, #3 (#2)}

\newcommand\MPLA[3]{\textsl{Mod. Phys. Lett.} \textbf{A#1}, #3 (#2)}

\newcommand\xdot{\stackrel{.}{x}}

\newcommand\etadot{\stackrel{.}{\eta}}

\begin{document}

\markboth{E. Guendelman, A. Kaganovich, E. Nissimov and S. Pacheva}
{Hiding and Confining Charges via ``Tube-like'' Wormholes}

%
\catchline{}{}{}{}{}
%

\title{HIDING AND CONFINING CHARGES VIA ``TUBE-LIKE'' WORMHOLES}

\author{Eduardo Guendelman and Alexander Kaganovich}

\address{Department of Physics, Ben-Gurion University of the Negev\\
P.O.Box 653, IL-84105 ~Beer-Sheva, Israel \\
guendel@bgu.ac.il, alexk@bgu.ac.il}

\author{Emil Nissimov and Svetlana Pacheva}

\address{Institute for Nuclear Research and Nuclear Energy, Bulgarian Academy
of Sciences\\
Boul. Tsarigradsko Chausee 72, BG-1784 ~Sofia, Bulgaria \\
nissimov@inrne.bas.bg, svetlana@inrne.bas.bg}

\maketitle

\begin{history}
\received{}
\revised{}
\end{history}

\vspace{-.1in}
\begin{abstract}
We describe two interesting effects in wormhole physics. First, we find
that a genuinely charged matter source of gravity and electromagnetism
may appear {\em electrically neutral} to an external observer -- a
phenomenon opposite to the famous Misner-Wheeler ``charge without charge''
effect. We show that this phenomenon takes place when coupling a bulk 
gravity/nonlinear-gauge-field system self-consistently to a codimension-one 
charged {\em lightlike} brane as a matter source. The ``charge-hiding'' effect 
occurs in a self-consistent wormhole
solution of the above coupled gravity/nonlinear-gauge-field/lightlike-brane system
which connects a non-compact ``universe'', comprising the exterior region of
Schwarzschild-(anti-)de-Sitter (or purely Schwarzschild) black hole beyond the 
internal (Schwarzschild) horizon, 
to a Levi-Civita-Bertotti-Robinson-type (``tube-like'') ``universe'' with two
compactified dimensions via a wormhole ``throat'' occupied by the charged
lightlike brane. In this solution the whole electric flux produced by the charged
lightlike brane is expelled into the compactified Levi-Civita-Bertotti-Robinson-type
``universe'' and, consequently, the brane is detected as neutral by an
observer in the Schwarzschild-(anti-)de-Sitter ``universe''.
Next, the above ``charge-hiding'' solution can be further generalized to a truly
{\em charge-confining} wormhole solution when we couple the bulk
gravity/nonlinear-gauge-field system self-consistently to {\em two} separate
codimension-one charged {\em lightlike} branes with equal in magnitude but opposite 
charges. The latter system possesses a ``two-throat'' wormhole solution, where the 
``left-most'' and the ``right-most'' ``universes''
are two identical copies of the exterior region of the neutral Schwarzschild-de-Sitter 
black hole beyond the Schwarzschild horizon,
whereas the ``middle'' ``universe'' is of generalized Levi-Civita-Bertotti-Robinson
``tube-like'' form with geometry $dS_2 \times S^2$ ($dS_2$ being the
two-dimensional de Sitter space). It comprises the
finite-extent intermediate region of $dS_2$ between its two horizons. Both
``throats'' are occupied by the two oppositely charged lightlike branes and the whole
electric flux produced by the latter is confined entirely within the middle
finite-extent ``tube-like'' ``universe''.
A crucial ingredient is the special form of the nonlinear gauge field
action, which contains both the standard Maxwell term as well as a square
root of the latter. This theory was previously shown to produce
a QCD-like confining dynamics in flat space-time.
\end{abstract}


\ccode{PACS numbers: 11.25.-w, 04.70.-s,04.50.+h}

\section{Introduction}

In Ref.~\refcite{tHooft-03} `t Hooft has proposed a consistent quantum description
of linear confinement phenomena in terms of effective nonlinear
gauge field actions, where the nonlinear terms play the role of effective 
``infrared counterterms''. In particular, he has argued that the energy density of
electrostatic field configurations should be a linear function of the
electric displacement field in the infrared region (see especially Eq.(5.10) in
Ref.~\refcite{tHooft-03}), which means that an additional term of the form of
square root of the standard Maxwell term
should appear in the effective action. The simplest way to realize this idea
in Minkowski space-time is by considering the following
nonlinear effective gauge field model \ct{GG-1}${}^{-}$\ct{GG-6}:
\br
S = \int d^4 x\, L(F^2) \quad ,\quad
L(F^2) = -\frac{1}{4} F^2 - \frac{f}{2} \sqrt{-F^2} \; ,
\lab{GG} \\
F^2 \equiv F_{\m\n} F^{\m\n} \quad ,\quad 
F_{\m\n} = \pa_\m A_\n - \pa_\n A_\m  \; ,
\nonu
\er
with $f$ being a positive coupling constant. 

Since the Lagrangian $L(F^2)$ in \rf{GG} contains both the 
usual Maxwell term as well as a non-analytic function of $F^2$, it is thus
a {\em non-standard} form of nonlinear electrodynamics. There are various reasons 
supporting the natural appearance of the ``square-root'' Maxwell term in effective
gauge field actions besides `t Hooft's arguments in Ref.~\refcite{tHooft-03}. 
Originally a purely ``square-root'' Lagrangian in flat space-time 
$- \frac{f}{2}\sqrt{F^2}$ (in ``magnetic''-dominated form) was proposed by 
Nielsen and Olesen \ct{N-O-1} to describe string dynamics
(see also Refs.~\refcite{N-O-2}-\refcite{N-O-4}). 
Furthermore, it has been shown in Refs.~\refcite{GG-1}-\refcite{GG-3} that the 
square root of the Maxwell term naturally arises (in flat space-time) as a result of 
spontaneous breakdown of scale symmetry of the original scale-invariant Maxwell 
theory with $f$ appearing as an integration constant responsible for the latter 
spontaneous breakdown.

As shown in Refs.~\refcite{GG-2}--\refcite{GG-6} the flat space-time model \rf{GG},
when coupled to charged fermions,
produces a confining effective potential $V(r) = - \frac{\a}{r} + \b r$ (Coulomb plus 
linear one) which is of the form of the well-known ``Cornell'' potential 
\ct{cornell-potential-1}${}^{-}$\ct{cornell-potential-3} in quantum chromodynamics (QCD).
Also, for static field configurations the model \rf{GG} yields the following electric 
displacement field $\vec{D} = \vec{E} - \frac{f}{\sqrt{2}}\frac{\vec{E}}{|\vec{E}|}$. 
The pertinent energy density turns out to be (there is {\em no} contribution from 
the square-root term in \rf{GG}) 
$\h \vec{E}^2 = \h |\vec{D}|^2 + \frac{f}{\sqrt{2}}|\vec{D}| + \frac{1}{4} f^2$, 
so that it indeed contains a term linear w.r.t. $|\vec{D}|$ as argued by `t Hooft
\ct{tHooft-03}. Similar connection between $\vec{D}$ and $\vec{E}$ has
been considered as an example of a ``classical model of confinement'' in
Ref.~\refcite{lehmann} and analyzed generalizing the methods
developed for the ``leading logarithm model'' in Ref.~\refcite{adler-piran}.

The gauge-field system with a square root of the Maxwell term \rf{GG}
coupled to gravity (cf. Eq.\rf{gravity+GG} below)
was recently studied in Ref.~\refcite{grav-cornell} (see the brief review in the
following Section 2), where the following interesting new
features of the pertinent static spherically symmetric solutions have been found:

(i) appearance of a constant vacuum radial electric field (in addition to the 
Coulomb one) in charged black holes within
Reissner-Nordstr{\"o}m-de-Sitter-type 
and/or Reissner-Nordstr{\"o}m-{\em anti}-de-Sitter-type 
space-times, in particular, in electrically neutral black holes with 
Schwarzschild-de-Sitter 
and/or Schwarzschild-{\em anti}-de-Sitter 
geometry;

(ii) novel mechanism of {\em dynamical generation} of cosmological constant
through the nonlinear gauge field dynamics due to the ``square-root'' Maxwell term;

(iii) appearance of a confining-type effective potential in charged test particle 
dynamics in the above black hole backgrounds.


Further, it is interesting to study possible new effects which can take
place in the context of wormhole physics where the wormholes are generated
due to the presence of nonlinear gauge fields with confining type dynamics.
To this end let us recall that
Misner-Wheeler ``charge without charge'' effect \ct{misner-wheeler} stands out 
as one of the most interesting physical phenomena produced by wormholes.
Misner and Wheeler realized that wormholes connecting two asymptotically flat 
space-times provide the possibility of existence of electromagnetically
non-trivial solutions, where the lines of force of the electric field flow from one 
universe to the other without a source and giving the impression of being 
positively charged in one universe and negatively charged in the other universe.

For a detailed exposition of the basics of wormhole physics we refer to
Visser's book Ref.~\refcite{visser-book+} (see also 
Refs.~\refcite{visser-1,visser-2}) and some more recent accounts 
\ct{WH-rev-1}${}^{-}$\ct{bronnikov-2}.

In a recent note \ct{hiding} we found the opposite effect in wormhole physics,
namely, that a genuinely charged matter source of gravity and electromagnetism
may appear {\em electrically neutral} to an external observer. We 
showed this phenomenon to take place in the coupled gravity/nonlinear-gauge-field
system \rf{gravity+GG} (without bare cosmological constant) self-consistently 
interacting with a charged lightlike brane as a matter source 
(cf. Eq.\rf{gravity+GG+LL} below).
In this case the lightlike brane, which connects as a wormhole ``throat'' 
a non-compact ``universe'' with a compactified ``universe'', is electrically charged, 
however all of its flux flows into the compactified (``tube-like'') ``universe'' only. 
No Coulomb field is produced in the non-compact ``universe'', therefore, the 
wormhole hides the charge from an external observer in the latter ``universe''.
This charge-hiding effect is exclusively due to the presence of the
``square-root'' Maxwell term in the nonlinear gauge field action.

A few remarks about the relevance of lightlike branes within the present
context are in order.
In our previous papers \ct{LL-main-1}${}^{-}$\ct{BR-WH-2} we have 
provided an explicit reparametrization invariant world-volume Lagrangian formulation of
lightlike $p$-branes (``\textsl{LL-branes}'' for short)
(a brief review is given in Section 4)
and we have used them to construct various types of wormhole, regular black hole and 
lightlike braneworld solutions in $D\!=\!4$ or 
higher-dimensional asymptotically flat or asymptotically (anti-)de Sitter bulk 
space-times. In particular, in Refs.~\refcite{BR-WH-1,BR-WH-2} we have
shown that \textsl{LL-branes} can trigger a series of spontaneous
compactification-decompactification transitions of space-time regions,
\textsl{e.g.}, from ordinary compactified (``tube-like'') 
Levi-Civita-Bertotti-Robinson space \ct{LC-BR-1}${}^{-}$\ct{LC-BR-3} 
to non-compact Reissner-Nordstr{\"o}m
or Reissner-Nordstr{\"o}m-de-Sitter region or {\sl vice versa}. 
Wormholes with ``tube-like'' structure and regular black holes with ``tube-like''
core have been previously obtained within different contexts in 
Refs.~\refcite{eduardo-dzhunu-1}-\refcite{zaslavskii-3}.

Here an important remark about ``Einstein-Rosen bridge'' wormhole is
in order. The nomenclature of ``Einstein-Rosen bridge'' in several standard textbooks 
uses the Kruskal-Szekeres manifold, where the ``Einstein-Rosen bridge'' geometry becomes
{\em dynamical} (see Ref.~\refcite{MTW}, p.839, Fig. 31.6, and Ref.~\refcite{Carroll},
p.228, Fig. 5.15). The latter notion of ``Einstein-Rosen bridge''
is {\em not} equivalent to the original Einstein-Rosen's construction 
\ct{einstein-rosen}, where the space-time manifold is {\em static} spherically
symmetric consisting of two identical
copies of the outer Schwarzschild space-time region ($r>2m$) glued together
along the horizon at $r=2m$. Namely, the two regions in Kruskal-Szekeres space-time 
corresponding to the outer Schwarzschild space-time region ($r>2m$) and labeled 
$(I)$ and $(III)$ in Ref.~\refcite{MTW} are generally
{\em disconnected} and share only a two-sphere (the angular part) as a common border
($U=0, V=0$ in Kruskal-Szekeres coordinates), whereas in the original Einstein-Rosen
``bridge'' construction \ct{einstein-rosen} the boundary between the two identical 
copies of the outer Schwarzschild space-time region ($r>2m$) is a three-dimensional 
(lightlike) hypersurface ($r=2m)$. In Refs.~\refcite{LL-main-4,Kerr-rot-WH-2} it
has been shown that the ``Einstein-Rosen bridge'' in its original formulation
\ct{einstein-rosen} naturally arises as the simplest particular case of static 
spherically symmetric wormhole solutions produced by lightlike branes as
gravitational sources, where the two identical ``universes'' with Schwarzschild
outer-region geometry are glued together by a lightlike brane occupying
their common horizon -- the wormhole ``throat''. An understanding of this
picture within the framework of Kruskal-Szekeres manifold was subsequently
provided in Ref.~\refcite{poplawski}, which involves Rindler's elliptic
identification of the two antipodal future event horizons.

Let us recall that \textsl{LL-branes} by themselves play an important role in modern
general relativity They are singular null (lightlike) 
hypersurfaces in Riemannian space-time which provide dynamical description of 
various physically important  phenomena in cosmology and astrophysics such as:
(i) impulsive lightlike signals arising in cataclysmic astrophysical events 
\ct{barrabes-hogan}; (ii) the ``membrane paradigm'' \ct{membrane-paradigm} of black 
hole physics; (iii) the thin-wall approach to domain walls coupled to 
gravity \ct{Israel-66}${}^{-}$\ct{Berezin-etal}.
More recently, \textsl{LL-branes} became significant also in the context of
modern non-perturbative string theory, in particular, as the so called
$H$-branes describing quantum horizons (black hole and cosmological)
\ct{kogan-01}, as Penrose limits of baryonic $D$-branes \ct{mateos-02}, \textsl{etc} 
(see also Refs.~\refcite{nonperturb-string-1,nonperturb-string-2}).

In the pioneering papers \ct{Israel-66}${}^{-}$\ct{Berezin-etal} \textsl{LL-branes}
in the context of gravity and cosmology have been extensively studied from a 
phenomenological point of view, \textsl{i.e.}, by introducing them without specifying
the Lagrangian dynamics from which they may originate\foot{In a more recent paper 
\ct{barrabes-israel-05} brane actions in terms of their pertinent extrinsic geometry
have been proposed which generically describe non-lightlike branes, whereas the 
lightlike branes are treated as a limiting case.}. 
On the other hand, we have proposed in a series of recent papers 
\ct{LL-main-1}${}^{-}$\ct{BR-WH-2} a new class of concise reparametrization 
invariant world-volume Lagrangian actions (see also Section 4 below), 
providing a derivation from first principles of the \textsl{LL-brane} dynamics.
The latter feature -- the explicit world-volume Lagrangian description of 
\textsl{LL-branes} is the principal distinction of our wormhole construction
via (charged) \textsl{LL-branes} as sources of gravity and electromagnetism
w.r.t. other non-Lagrangian ``thin-shell'' constructions of wormhole solutions
(for the basics of the ``thin-shell'' cut-and-paste technique we refer to
the book Ref.~\refcite{visser-book+}). 

There are several characteristic features of \textsl{LL-branes} which drastically
distinguish them from ordinary Nambu-Goto branes: 

(i) They describe intrinsically lightlike modes, whereas Nambu-Goto branes describe
massive ones.

(ii) The tension of the \textsl{LL-brane} arises as an {\em additional
dynamical degree of freedom}, whereas Nambu-Goto brane tension is a given
{\em ad hoc} constant. 
The latter characteristic feature significantly distinguishes our \textsl{LL-brane}
models from the previously proposed {\em tensionless} $p$-branes (for a
review of the latter, see Ref.~\refcite{lindstroem-etal}) which rather resemble 
a $p$-dimensional continuous distribution of massless point-particles. 

(iii) Consistency of \textsl{LL-brane} dynamics in a spherically or axially
symmetric gravitational background of codimension one requires the presence
of a horizon which is automatically occupied by the \textsl{LL-brane}
(``horizon straddling'' according to the terminology of 
Ref.~\refcite{Barrabes-Israel}).

(iv) When the \textsl{LL-brane} moves as a {\em test} brane in spherically or 
axially symmetric gravitational backgrounds its dynamical tension exhibits 
exponential ``inflation/deflation'' time behavior \ct{inflation-all-1}
-- an effect similar to the ``mass inflation'' effect around black hole horizons
\ct{israel-poisson-1,israel-poisson-2}. 

The principal object of study in the present paper is the self-consistently 
coupled gravity/nonlinear-gauge-field system, containing the square root of
the Maxwell term, with one {\em or more} \textsl{LL-brane(s)}. We significantly 
extend the results of our previous note \ct{hiding} by constructing both more 
general wormhole solutions displaying a ``charge-hiding'' effect as well as
completely new ``two-throat'' wormhole solution with genuinely QCD-like {\em
confining} behavior.

The plan of the present paper is as follows. 
In Section 2 we briefly review the Lagrangian formulation and the
corresponding static spherically symmetric solutions of the coupled 
gravity/nonlinear-gauge-field system \rf{gravity+GG} \ct{grav-cornell}, including the
generation of vacuum constant-magnitude electric field as well as dynamical 
generation of cosmological constant.

In Section 3 we extend the results of Ref.~\refcite{hiding} obtaining new
solutions of compactified Levi-Civita-Bertotti-Robinson type in the
gravity/nonlinear-gauge-field system \rf{gravity+GG} depending on
the magnitude of the bare cosmological constant versus the dynamically
generated one.

In Section 4 we briefly review the world-volume Lagrangian formulation and
the basic properties of \textsl{LL-brane} dynamics, particularly stressing
on the ``horizon straddling'' property (cf. Eqs.\rf{straddling} below).

In Section 5 we describe the Lagrangian formulation of the self-consistently
coupled bulk gravity/nonlinear-gauge-field system \rf{gravity+GG} to one or more
\textsl{LL-brane} sources (cf. Eq.\rf{gravity+GG+LL}) and outline a general
procedure to derive wormhole solutions.
 
In Section 6 we construct ``one-throat'' wormhole solutions to the coupled
gravity/nonlinear-gauge-field/\textsl{LL-brane} system \rf{gravity+GG+LL} with
the charged \textsl{LL-brane} occupying the wormhole ``throat'', which connects
a non-compact ``universe'' with Schwarzschild-(anti)-de-Sitter geometry 
(where the cosmological constant is partially or entirely {\em dynamically} 
generated) to a compactified (``tube-like'') ``universe'' of 
Levi-Civita-Bertotti-Robinson type. These wormholes exhibit
the novel property of {\em hiding} electric charge from external observer in the
non-compact ``universe'', \textsl{i.e.}, the whole electric flux produced by
the charged \textsl{LL-brane} at the wormhole ``throat'' is expelled into the 
``tube-like'' ``universe''.

In Section 7 we construct more general ``two-throat'' wormhole solution to the 
coupled gravity/nonlinear-gauge-field/\textsl{LL-brane} system \rf{gravity+GG+LL}
with two separate charged \textsl{LL-branes} with equal in magnitude but opposite 
charges occupying the wormhole ``throats'' and connecting pairwise three different
``universes''. The ``left-most'' and the ``right-most'' ``universes''
are two identical copies of the exterior region of the electrically neutral 
Schwarzschild-de-Sitter black hole beyond the Schwarzschild horizon.
The ``middle'' ``universe'' is of Levi-Civita-Bertotti-Robinson
``tube-like'' form with geometry $dS_2 \times S^2$ ($dS_2$ being the
two-dimensional de Sitter space). It comprises the finite-extent intermediate 
region of $dS_2$ between its two horizons. Both oppositely charged 
\textsl{LL-branes} occupying the two ``throats'' are producing an electric
flux which turns out to be {\em confined} entirely within the middle
finite-extent ``tube-like'' ``universe'', \textsl{i.e.}, no flux from the
charged \textsl{LL-branes} is flowing into the non-compact outer ``universes''.


\section{Gravity/Nonlinear-Gauge-Field System. Spherically Symmetric Solutions}

We will consider the simplest coupling to gravity of the nonlinear gauge field system
with a square root of the Maxwell term \rf{GG} known to produce QCD-like confinement
in flat space-time \ct{GG-2}${}^{-}$\ct{GG-6}. 
The relevant action is given by (we use units with Newton constant $G_N=1$):
\br
S = \int d^4 x \sqrt{-G} \Bigl\lb \frac{R(G) - 2\L}{16\pi} + L(F^2)\Bigr\rb 
\quad ,\quad
L(F^2) = - \frac{1}{4} F^2 - \frac{f}{2} \sqrt{\vareps F^2} \; ,
\lab{gravity+GG} \\
F^2 \equiv F_{\k\l} F_{\m\n} G^{\k\m} G^{\l\n} \quad ,\quad 
F_{\m\n} = \pa_\m A_\n - \pa_\n A_\m \; .
\nonu
\er
Here $R(G)$ is the scalar curvature of the space-time metric
$G_{\m\n}$ and $G \equiv \det\Vert G_{\m\n}\Vert$; the sign factor $\vareps = \pm 1$
in the square-root term in \rf{gravity+GG} corresponds to ``magnetic'' or ``electric''
dominance; $f$ is a positive coupling constant. 

It is important to stress that we {\em do not} need to introduce {\em any} bare 
cosmological constant $\L$ in \rf{gravity+GG} since the ``square-root'' Maxwell term 
dynamically generates a {\em non-zero effective cosmological constant} 
$\L_{\rm{eff}} = 2\pi f^2$ \ct{hiding}. The role of the bare $\L$ is just shifting
the effective $\L_{\rm{eff}}$ (see Eqs.\rf{CC-eff} below).

\vspace{.1in}
\textbf{Remark.} One could start with the non-Abelian version of the 
gauge field action in \rf{gravity+GG}. Since we will be interested in static 
spherically symmetric solutions, the non-Abelian gauge theory effectively reduces 
to an Abelian one as pointed out in Ref.~\refcite{GG-2}.

\vspace{.1in}
The corresponding equations of motion read:
\be
R_{\m\n} - \h G_{\m\n} R + \L G_{\m\n} = 8\pi T^{(F)}_{\m\n} \; ,
\lab{einstein-eqs}
\ee
\be
T^{(F)}_{\m\n} = \Bigl( 1 + 
\vareps \frac{f}{\sqrt{\vareps F^2}}\Bigr) F_{\m\k} F_{\n\l} G^{\k\l}
- \frac{1}{4} \Bigl( F^2 + 2f\sqrt{\vareps F^2}\Bigr) G_{\m\n} \; ,
\lab{stress-tensor-F}
\ee
and
\be
\pa_\n \(\sqrt{-G}\Bigl( 1 +
\vareps\frac{f}{\sqrt{\vareps F^2}}\Bigr) F_{\k\l} G^{\m\k} G^{\n\l}\)=0
\; .
\lab{GG-eqs}
\ee

Here we will first consider the case of ``electric dominance'', \textsl{i.e.},
$\vareps = -1$ in \rf{gravity+GG}.
In our preceding paper \ct{grav-cornell} we have shown that the gravity-gauge-field
system \rf{gravity+GG} with {\em zero} bare cosmological constant possesses static 
spherically symmetric solutions with a radial electric field containing {\em both}
Coulomb and {\em constant} ``vacuum'' pieces:
\be
F_{0r} = \frac{\vareps_F f}{\sqrt{2}} + \frac{Q}{\sqrt{4\pi}\, r^2} 
\quad ,\quad \vareps_F \equiv \mathrm{sign}(F_{0r}) = \mathrm{sign}(Q) \; ,
\lab{cornell-sol}
\ee
and the space-time metric:  
\br
ds^2 = - A(r) dt^2 + \frac{dr^2}{A(r)} + r^2 \bigl(d\th^2 + \sin^2 \th d\vp^2\bigr)
\; ,
\lab{spherical-static} \\
A(r) = 1 - \sqrt{8\pi}|Q|f - \frac{2m}{r} + \frac{Q^2}{r^2} - \frac{2\pi f^2}{3} r^2 \; ,
\lab{RN-dS+const-electr}
\er
is Reissner-Nordstr{\"o}m-de-Sitter-type with {\em dynamically generated} effective 
cosmological constant $2\pi f^2$. 
In the presence of the bare cosmological constant term in \rf{gravity+GG} 
the only effect is shifting of the effective cosmological constant, namely:
\be
A(r) = 1 - \sqrt{8\pi}|Q|f - \frac{2m}{r} + \frac{Q^2}{r^2} - \frac{\L_{\mathrm{eff}}}{3} r^2 
\quad ,\quad \L_{\mathrm{eff}} = 2\pi f^2 + \L \; .
\lab{CC-eff}
\ee
The expression for $\L_{\mathrm{eff}}$ \rf{CC-eff} tells us that:
\begin{itemize}
\item
Solution \rf{cornell-sol}--\rf{spherical-static} with \rf{CC-eff} is 
Reissner-Nordstr{\"o}m-de-Sitter-type with additional constant vacuum radial electric 
field even for {\em negative} bare cosmological constant $\L < 0$ provided 
$|\L| < 2\pi f^2$, \textsl{i.e.}, $\L_{\mathrm{eff}}>0$ in \rf{CC-eff};
\item
Solution \rf{cornell-sol}--\rf{spherical-static} with \rf{CC-eff} becomes
Reissner-Nordstr{\"o}m-type with additional constant vacuum radial 
electric field in spite of the presence of {\em negative} bare cosmological constant
$\L < 0$ with $|\L| = 2\pi f^2$, \textsl{i.e.}, $\L_{\mathrm{eff}}=0$ in \rf{CC-eff};
\item
Solution \rf{cornell-sol}--\rf{spherical-static} with \rf{CC-eff} is 
Reissner-Nordstr{\"o}m-{\em anti}-de-Sitter-type with constant vacuum radial electric 
field for sufficiently large {\em negative} bare cosmological constant $\L < 0$ 
with $|\L| > 2\pi f^2$, \textsl{i.e.}, $\L_{\mathrm{eff}}<0$ in \rf{CC-eff}.
\end{itemize}

Notice that the ``leading'' constant term in the Reissner-Nordstr{\"o}m-(anti-)de-Sitter-type 
metric coefficient \rf{CC-eff} is different from 1 when $Q \neq 0$. This effect resembles 
the effect on gravity produced by a spherically symmetric ``hedgehog'' configuration 
of a nonlinear sigma-model scalar field with $SO(3)$ symmetry
(see Refs. ~\refcite{Ed-Rab-hedge,barriola-vilenkin}).

The electrically neutral case $Q=0$ will play an important role in what
follows:
\be
A(r) = 1 - \frac{2m}{r} - \frac{\L_{\mathrm{eff}}}{3} r^2 \;\; ,\;\;
\L_{\mathrm{eff}} = 2\pi f^2 + \L \quad ,\quad 
F_{0r} = \frac{\vareps_F f}{\sqrt{2}} \; .
\lab{S-dS+const-electr}
\ee

\begin{itemize}
\item
Solution \rf{S-dS+const-electr} is Schwarzschild-de-Sitter black hole carrying a 
constant vacuum radial electric field for all $\L > - 2\pi f^2$, even for
negative $\L$ provided $|\L| < 2\pi f^2$, \textsl{i.e.}, $\L_{\mathrm{eff}}>0$ 
in \rf{S-dS+const-electr};
\item
Solution \rf{S-dS+const-electr} for negative $\L$ with $|\L| = 2\pi f^2$ becomes
asymptotically flat ordinary Schwarzschild carrying a constant vacuum radial 
electric field in spite of the presence of {\em negative} bare cosmological constant
\textsl{i.e.}, $\L_{\mathrm{eff}}=0$ in \rf{S-dS+const-electr};
\item
Solution \rf{S-dS+const-electr} is Schwarzschild-{\em anti}-de-Sitter carrying a 
constant vacuum radial electric field for all $\L < 0$ with $|\L| > 2\pi f^2$,
\textsl{i.e.}, $\L_{\mathrm{eff}}<0$ in \rf{S-dS+const-electr}.
\end{itemize}

\section{Generalized Levi-Civita-Bertotti-Robinson Space-Times}
Here we will look for static solutions of Levi-Civita-Bertotti-Robinson type 
\ct{LC-BR-1}${}^{-}$\ct{LC-BR-3} of the system \rf{einstein-eqs}--\rf{GG-eqs}, namely, 
with space-time geometry of the form $\cM_2 \times S^2$, where $\cM_2$ is some 
two-dimensional manifold:
\be
ds^2 = - A(\eta) dt^2 + \frac{d\eta^2}{A(\eta)} 
+ r_0^2 \bigl(d\th^2 + \sin^2 \th d\vp^2\bigr) \;\; ,\;\;  
-\infty < \eta <\infty \;\; ,\;\; r_0 = \mathrm{const}
\; ,
\lab{gen-BR-metric}
\ee
and being:
\begin{itemize}
\item
either purely electric type, where the sign factor $\vareps = -1$ in the gauge 
field Lagrangian $L(F^2)$ \rf{gravity+GG}:
\be
F_{\m\n} = 0 \;\; \mathrm{for}\; \m,\n\neq 0,\eta \quad ,\quad
F_{0\eta} = F_{0\eta} (\eta) \; ;
\lab{electr-static}
\ee
\item
or purely magnetic type, where $\vareps = +1$ in \rf{gravity+GG}:
\be
F_{\m\n} = 0 \;\; \mathrm{for}\; \m,\n\neq i,j\equiv \th,\vp \quad ,\quad
\pa_0 F_{ij} = \pa_\vp F_{ij} = 0 \; .
\lab{magnet-static}
\ee
\end{itemize}

In the purely electric case \rf{electr-static} the gauge field equations of
motion become:
\be
\pa_\eta \Bigl( F_{0\eta} - \frac{\vareps_F f}{\sqrt{2}}\Bigr) = 0
\quad ,\quad \vareps_F \equiv \mathrm{sign}(F_{0\eta}) \; ,
\lab{GG-eqs-0}
\ee
yielding a globally constant vacuum electric field:
\be
F_{0\eta} = c_F = \mathrm{arbitrary ~const} \; .
\lab{const-electr}
\ee
The (mixed) components of energy-momentum tensor \rf{stress-tensor-F} read:
\be
{T^{(F)}}^0_0 = {T^{(F)}}^\eta_\eta = - \h F^2_{0\eta} \quad ,\quad
T^{(F)}_{ij} = g_{ij}\Bigl(\h F^2_{0\eta} - \frac{f}{\sqrt{2}}|F_{0\eta}|\Bigr)
\; .
\lab{T-F-electr}
\ee
Taking into account \rf{T-F-electr}, the Einstein eqs.\rf{einstein-eqs} for
$(ij)$, where $R_{ij}=\frac{1}{r_0^2} g_{ij}$ because of the $S^2$ factor in
\rf{gen-BR-metric}, yield:
\be
\frac{1}{r_0^2} = 4\pi c_F^2 + \L \; .
\lab{einstein-ij}
\ee
The $(00)$ Einstein eq.\rf{einstein-eqs} using the
expression $R^0_0 = - \h \pa_\eta^2 A$ (valid for metrics of the type
\rf{gen-BR-metric}, cf. Ref.~\refcite{Ed-Rab-1,Ed-Rab-2}) becomes:
\be
\pa_\eta^2 A = 8\pi h(|c_F|) \quad ,\quad
h(|c_F|) \equiv c_F^2 - \sqrt{2}f|c_F| - \frac{\L}{4\pi} \; ,
\lab{einstein-00}
\ee

Thus, we arrive at the following three distinct types of
Levi-Civita-Bertotti-Robinson solutions for gravity coupled to the
non-linear gauge field system \rf{gravity+GG}:

(i) $AdS_2 \times S^2$ with constant vacuum electric field
$|F_{0\eta}| = |c_F|$, where $AdS_2$ is two-dimensional anti-de Sitter 
space with (using the definition of $h(|c_F|)$ in \rf{einstein-00}):
\be
A(\eta) = 4\pi h(|c_F|)\,\eta^2 \quad ,\quad h(|c_F|) >0 
\lab{AdS2}
\ee
in the metric \rf{gen-BR-metric}, $\eta$ being the Poincare patch
space-like coordinate, provided:
\be
|c_F| > \frac{f}{\sqrt{2}}\Bigl( 1 + \sqrt{1 + \frac{\L}{2\pi f^2}}\Bigr) 
\quad \mathrm{for}\;\; \L \geq - 2\pi f^2 \; ,
\lab{AdS2-cF-1}
\ee
\be
|c_F| > \sqrt{\frac{1}{4\pi}|\L|} 
\quad \mathrm{for}\;\; \L < 0 \;\; ,\;\; |\L| > 2\pi f^2 \; .
\lab{AdS2-cF-3}
\ee

(ii) $Rind_2 \times S^2$ with constant vacuum electric field 
$|F_{0\eta}| = |c_F|$, where $Rind_2$ is the flat two-dimensional 
Rindler space with:
\be
A(\eta) = \eta \;\; \mathrm{for}\; 0 < \eta < \infty \quad \mathrm{or} \quad
A(\eta) = - \eta \;\; \mathrm{for}\; -\infty <\eta < 0 
\lab{Rindler2}
\ee
in the metric \rf{gen-BR-metric}, provided:
\be
|c_F| = \frac{f}{\sqrt{2}}\Bigl( 1 + \sqrt{1 + \frac{\L}{2\pi f^2}}\Bigr) 
\quad \mathrm{for}\;\; \L \geq - 2\pi f^2 \; ,
\lab{Rindler2-cF-1}
\ee

(iii)  $dS_2 \times S^2$ with weak constant vacuum electric field
$|F_{0\eta}| = |c_F|$, where $dS_2$ is two-dimensional de Sitter space with:
\be
A(\eta) = 1 - K(|c_F|)\,\eta^2 \;\; ,\;\; 
K (|c_F|) \equiv - 4\pi h(|c_F|) \equiv 4\pi\Bigl(\sqrt{2}f |c_F| - c_F^2 +
\frac{\L}{4\pi}\Bigr) >0
\lab{dS2}
\ee
in the metric \rf{gen-BR-metric}, provided:
\be
|c_F| < \frac{f}{\sqrt{2}}\Bigl( 1 + \sqrt{1 + \frac{\L}{2\pi f^2}}\Bigr)
\quad \mathrm{for}\;\; \L > - 2\pi f^2 \; .
\lab{dS2-cF-1}
\ee
When $\L = 0$, for the special value $|c_F| = \frac{f}{\sqrt{2}}$ we recover the 
Nariai solution \ct{nariai-1,nariai-2} with $A(\eta) = 1 - 2\pi f^2 \eta^2$ and 
equality (up to signs) among energy density, radial and transverse pressures:
$\rho = - p_r = - p_{\perp} = \frac{f^2}{4}$ 
(using standard definitions 
${T^{(F)}}^\m_\n = \mathrm{diag} \(-\rho,p_r,p_{\perp},p_{\perp}\)$).

In all three cases above the size of the $S^2$ factor is given by \rf{einstein-ij}.
Solutions \rf{Rindler2} and \rf{dS2} with $\L=0$ are
specifically due to the presence of the non-Maxwell square-root term 
(with $\vareps =-1$) in the gauge field Lagrangian \rf{gravity+GG}.

In the purely magnetic case \rf{magnet-static} the gauge field equations of
motion \rf{GG-eqs}:
\be
\pa_\n \Bigl\lb \sin\th \Bigl( 1 + \frac{f}{\sqrt{F^2}}\Bigr) F^{\m\n}\Bigr\rb = 0
\lab{GG-eqs-1}
\ee
yield magnetic monopole solution: 
\be
F_{ij} = B r_0^2\sin\th\,\vareps_{ij} \;\; ,\;\; B=\mathrm{const} \; , 
\lab{monopole}
\ee
irrespective of the presence of the ``square-root'' Maxwell term. The latter, 
however, does contribute to the energy-momentum tensor:
\be
{T^{(F)}}^0_0 = {T^{(F)}}^\eta_\eta = - \h B^2 - f|B| \quad ,\quad
T^{(F)}_{ij} = \h g_{ij} B^2 \; .
\lab{T-F-magnet}
\ee
Taking into account \rf{T-F-magnet}, the Einstein eqs.\rf{einstein-eqs} for $(ij)$
yield (\textsl{cf.} \rf{einstein-ij}):
\be
\frac{1}{r_0^2} = 4\pi\( B^2 + \sqrt{2}f|B|\) + \L \; ,
\lab{einstein-ij-1}
\ee
which determines the size of the $S^2$ factor,
whereas the mixed-component $(00)$ Einstein eq.\rf{einstein-eqs} gives:
\be
\pa^2_\eta A = 8\pi B^2 - 2\L \; .
\lab{einstein-00-1}
\ee

Thus, in the purely magnetic case we recover the three types of
Levi-Civita-Bertotti-Robinson solutions with constant-magnitude magnetic field:\\
%
(a) $AdS_2 \times S^2$ space-time with magnetic monopole \rf{monopole}
for $\L < 4\pi B^2$ with 
$A(\eta) = 4\pi \Bigl( B^2 - \frac{\L}{4\pi}\Bigr) \eta^2$ 
in the metric \rf{gen-BR-metric};\\
(b) $Rind_2 \times S^2$ space-time with magnetic monopole \rf{monopole}
for $\L = 4\pi B^2$ with
$A(\eta) = \eta \;\; ,\; \eta >0$ in the metric \rf{gen-BR-metric};\\
(c) $dS_2 \times S^2$ space-time with magnetic monopole \rf{monopole}
for $\L > 4\pi B^2$ with
$A(\eta) = 1 - 4\pi \Bigl(\frac{\L}{4\pi} - B^2\Bigr) \eta^2$ 
in the metric \rf{gen-BR-metric}.\\
Here the only feature is the dependence of the size of the $S^2$-factor on the 
``square-root'' Maxwell coupling constant $f$ \rf{einstein-ij-1}.

Generalized Levi-Civita-Bertotti-Robinson solutions of the above type have
already appeared in different contexts in 
Refs.~\refcite{zaslavskii-mat}-\refcite{zaslavskii-3} and Ref.~\refcite{cardoso-lemos-etal}
(extension to higher space-time dimensions).
The main distinction in the present case is that the Levi-Civita-Bertotti-Robinson
solutions are now generated 
due to the presence of the ``square-root'' Maxwell term in \rf{gravity+GG}
which also produces a non-zero effective cosmological constant.

In Ref.~\refcite{bronnikov-3} a different kind on nonlinear gauge field
Lagrangian $L(F^2)$ coupled to gravity has been considered which generates {\em locally} 
(in the vicinity of the center of the geometry) an effective cosmological constant. 
However, the latter $L(F^2)$
is an {\em analytic} function of $F^2$ reducing to the ordinary Maxwell term for
small $F^2$ unlike the present nonlinear Lagrangian $L(F^2)$ in \rf{gravity+GG}
containing the square-root term $\sqrt{-F^2}$. This latter feature of \rf{gravity+GG}
produces a {\em globally} defined dynamically generated cosmological constant $2\pi f^2$.

\section{Lagrangian Formulation of Lightlike Branes. Horizon ``Straddling''}
\label{LL-brane}

In what follows we will consider gravity/gauge-field system
self-consistently interacting with a lightlike $p$-brane (``\textsl{LL-brane}'' for short)
of codimension one ($D=(p+1)+1$); in the present Section will keep arbitrary
the number of space-time dimensions). In a series of previous papers 
\ct{LL-main-1}${}^{-}$\ct{BR-WH-2}
we have proposed manifestly reparametrization invariant world-volume Lagrangian 
formulation in several dynamically equivalent forms of \textsl{LL-branes} coupled to bulk 
gravity $G_{\m\n}$ and bulk gauge fields, in particular, electromagnetic field $A_\m$. 
Here we will use our Polyakov-type formulation given by the world-volume action
\ct{BR-WH-1,BR-WH-2}:
\br
S_{\rm LL}\lb q\rb  = - \h \int d^{p+1}\s\, T b_0^{\frac{p-1}{2}}\sqrt{-\g}
\llb \g^{ab} {\bar g}_{ab} - b_0 (p-1)\rrb \; ,
\lab{LL-action+EM} \\
{\bar g}_{ab} \equiv \pa_a X^\m G_{\m\n} \pa_b X^\n 
- \frac{1}{T^2} (\pa_a u + q\cA_a)(\pa_b u  + q\cA_b) 
\quad , \quad \cA_a \equiv \pa_a X^\m A_\m \; .
\lab{ind-metric-ext-A}
\er
Here and below the following notations are used:
\begin{itemize}
\item
$\g_{ab}$ is the {\em intrinsic} Riemannian metric on the world-volume with
$\g = \det \Vert\g_{ab}\Vert$;
$g_{ab}$ is the {\em induced} metric on the world-volume:
\be
g_{ab} \equiv \pa_a X^{\m} G_{\m\n}(X) \pa_b X^{\n} \; ,
\lab{ind-metric}
\ee
which becomes {\em singular} on-shell (manifestation of the lightlike nature, 
\textsl{cf.} Eq.\rf{on-shell-singular-A} below); 
$b_0$ is a positive constant measuring the world-volume ``cosmological constant''.
\item
$X^\m (\s)$ are the $p$-brane embedding coordinates in the bulk
$D$-dimensional space-time with Riemannian metric
$G_{\m\n}(x)$ ($\m,\n = 0,1,\ldots ,D-1$); 
$(\s)\equiv \(\s^0 \equiv \t,\s^i\)$ with $i=1,\ldots ,p$ ;
$\pa_a \equiv \partder{}{\s^a}$.
\item
$u$ is auxiliary world-volume scalar field defining the lightlike direction
of the induced metric 
(see Eq.\rf{on-shell-singular-A} below) and it is a non-propagating degree of freedom
\ct{BR-WH-2}.
\item
$T$ is {\em dynamical (variable)} brane tension (also a non-propagating
degree of freedom). 
\item
Coupling parameter $q$ is the surface charge density of the \textsl{LL-brane}.
\end{itemize}

The corresponding equations of motion w.r.t. $X^\m$, $u$, $\g_{ab}$ and $T$ read accordingly
(using short-hand notation \rf{ind-metric-ext-A}):
\br
\pa_a \( T \sqrt{|{\bar g}|} {\bar g}^{ab}\pa_b X^\m\)
+ T \sqrt{|{\bar g}|} {\bar g}^{ab} \pa_a X^\l \pa_b X^\n \G^\m_{\l\n}
\nonu \\
+ \frac{q}{T} \sqrt{|{\bar g}|} {\bar g}^{ab}
\pa_a X^\n (\pa_b u  + q\cA_b) \cF_{\l\n}G^{\m\l} = 0 \; ,
\lab{X-eqs-NG-A} \\
\pa_a \(\frac{1}{T} \sqrt{|{\bar g}|} {\bar g}^{ab}(\pa_b u  + q\cA_b)\) = 0
\quad ,\quad  
\g_{ab} = \frac{1}{b_0} {\bar g}_{ab}  \; ,
\lab{u-gamma-eqs-NG-A} \\
T^2 + {\bar g}^{ab}(\pa_a u  + q\cA_a)(\pa_b u  + q\cA_b) = 0 \; .
\lab{T-eq-NG-A}
\er
Here ${\bar g} = \det\Vert {\bar g}_{ab} \Vert$ and $\G^\m_{\l\n}$ denotes the Christoffel 
connection for the bulk metric $G_{\m\n}$.

The on-shell singularity of the induced metric $g_{ab}$ \rf{ind-metric}, \textsl{i.e.}, 
the lightlike property, directly follows from Eq.\rf{T-eq-NG-A} and the definition of
${\bar g}_{ab}$ \rf{ind-metric-ext-A}:
\be
g_{ab} \({\bar g}^{bc}(\pa_c u  + q\cA_c)\) = 0 \; .
\lab{on-shell-singular-A}
\ee

Explicit world-volume reparametrization invariance of the \textsl{LL-brane} action
\rf{LL-action+EM} allows to introduce the standard synchronous gauge-fixing conditions
for the intrinsic world-volume metric  
$\g_{00} = -1\; ,\; \g_{0i} = 0 \; (i=1,\ldots,p)$,
which reduces Eqs.\rf{u-gamma-eqs-NG-A}--\rf{T-eq-NG-A} to the following relations:
\br
\frac{(\pa_0 u + q\cA_0)^2}{T^2} = b_0 + g_{00} \quad ,\quad 
\pa_i u + q\cA_i= (\pa_0 u + q\cA_0) g_{0i} \( b_0 + g_{00}\)^{-1} \; ,
\nonu \\
g_{00} = g^{ij} g_{0i} g_{0j} \;\; ,\;\;
\pa_0 \(\sqrt{g^{(p)}}\) + \pa_i \(\sqrt{g^{(p)}}g^{ij} g_{0j}\) = 0 \;\; ,\;\; 
g^{(p)} \equiv \det\Vert g_{ij}\Vert \; ,
\lab{g-rel}
\er
(recall that $g_{00},g_{0i},g_{ij}$ are the components of the induced metric 
\rf{ind-metric}; $g^{ij}$ is the inverse matrix of $g_{ij}$).

In our previous papers \ct{LL-main-1}${}^{-}$\ct{BR-WH-2}
we have studied in some detail the consistency of \textsl{LL-brane} dynamics in static
``spherically-symmetric''-type backgrounds, whose generic form reads (in what follows we
will use Eddington-Finkelstein coordinates \ct{EF-coord-1,EF-coord-2} where
$dt=dv-\frac{d\eta}{A(\eta)}\; ,\; F_{0\eta} = F_{v\eta}$):
\be
ds^2 = - A(\eta) dv^2 + 2dv d\eta + C(\eta) h_{ij}(\th) d\th^i d\th^j \quad ,\quad
F_{v\eta} = F_{v\eta} (\eta)\; , 
\lab{static-spherical-EF}
\ee
all remaining components of $F_{\m\n}$ being zero.
For the \textsl{LL-brane} we use the standard embedding ansatz:
\be
X^0\equiv v = \t \quad, \quad X^1\equiv \eta = \eta (\t) \quad, \quad 
X^i\equiv \th^i = \s^i \;\; (i=1,\ldots ,p) \; .
\lab{X-embed}
\ee
For the class of backgrounds \rf{static-spherical-EF} with the embedding
\rf{X-embed} (where the induced metric components $g_{0i}=0$) Eqs.\rf{g-rel}
reduce to:
\be
g_{00} = 0 \;\; ,\;\; 
\pa_0 C \bigl(\eta (\t)\bigr) \equiv \etadot \pa_\eta C\bv_{\eta = \eta (\t)} = 0 
\;\; ,\;\;
\frac{(\pa_0 u + q\cA_0)^2}{T^2} = b_0 \;\; ,\;\; \pa_i u =0 
\lab{g-rel-0}
\ee
($\etadot \equiv \pa_0 \eta \equiv \pa_\t \eta (\t)$).
Thus, in the generic case of non-trivial dependence of $C(\eta)$ on the
``radial-like'' coordinate $\eta$, the first two relations in \rf{g-rel-0} yield:
\be
\etadot = \h A \bigl(\eta (\t)\bigr) \quad ,\quad \etadot = 0 \quad \to \quad
\eta (\t) = \eta_0 = \mathrm{const} \quad ,\quad A(\eta_0) = 0 \; .
\lab{straddling}
\ee
In other words, consistency of \textsl{LL-brane} dynamics requires the corresponding
background \rf{static-spherical-EF} to possess a horizon at some $\eta\!=\!\eta_0$,
which is {\em automatically occupied} by the \textsl{LL-brane}.

The latter property is called ``horizon straddling'' according to the terminology of 
Ref.~\refcite{Barrabes-Israel}. Similar ``horizon straddling'' has been found also for 
\textsl{LL-branes} moving in rotating axially symmetric (Kerr or Kerr-Newman) and 
rotating cylindrically symmetric black hole backgrounds 
\ct{Kerr-rot-WH-1,Kerr-rot-WH-2}. 

\section{Bulk Gravity/Nonlinear-Gauge-Field System Coupled to Lightlike Brane Sources}
We consider now bulk Einstein/non-linear gauge field system \rf{gravity+GG} 
self-consistently coupled to $N\geq 1$ distantly separated charged codimension-one
{\em lightlike} branes (in the present case $D=4\,,\, p=2$). 
The pertinent Lagrangian action reads:
\be
S = \int d^4 x \sqrt{-G} \Bigl\lb \frac{R(G) - 2\L}{16\pi} + L(F^2)\Bigr\rb 
+ \sum_{k=1}^N S_{\mathrm{LL}}\lb q^{(k)}\rb
\;\; ,\;\; L(F^2) = - \frac{1}{4} F^2 - \frac{f}{2} \sqrt{- F^2} \; ,
\lab{gravity+GG+LL}
\ee
where $S_{\mathrm{LL}}\lb q^{(k)}\rb$ indicates the world-volume action
of the $k$-th \textsl{LL-brane} of the form \rf{LL-action+EM}. Henceforth we
will consider the case of ``electric dominance'' for the ``square-root''
Maxwell term.

The corresponding equations of motion are as follows:
\br
R_{\m\n} - \h G_{\m\n} R + \L G_{\m\n} = 
8\pi \Bigl\lb T^{(F)}_{\m\n} + \sum_{k=1}^N T^{(brane-k)}_{\m\n}\Bigr\rb \; ,
\lab{einstein+LL-eqs} \\
\pa_\n \Bigl\lb\sqrt{-G} \Bigl( 1 - \frac{f}{\sqrt{-F^2}}\Bigr) 
F_{\k\l} G^{\m\k} G^{\n\l}\Bigr\rb + \sum_{k=1}^N j_{(\mathrm{brane-k})}^\m = 0 \; .
\lab{GG+LL-eqs}
\er
Here $T^{(F)}_{\m\n}$ is the same as in \rf{stress-tensor-F}, whereas 
the energy-momentum tensor and the charge current density of $k$-th 
\textsl{LL-brane} are straightforwardly derived from the pertinent \textsl{LL-brane} 
action \rf{LL-action+EM}:
\be
T_{(brane-k)}^{\m\n} = 
- \int\!\! d^3\s\,\frac{\d^{(4)}\Bigl(x-X_{(k)}(\s)\Bigr)}{\sqrt{-G}}
\, T^{(k)}\,\sqrt{|{\bar g}_{(k)}|} {\bar g}_{(k)}^{ab}
\pa_a X_{(k)}^\m \pa_b X_{(k)}^\n \; ,
\lab{T-brane-A}
\ee
\br
j_{(\mathrm{brane-k})}^\m = -
q^{(k)} \int\!\! d^3\s\,\d^{(4)}\Bigl(x-X_{(k)}(\s)\Bigr)
\sqrt{|{\bar g}_{(k)}|} {\bar g}_{(k)}^{ab}\pa_a X_{(k)}^\m 
\frac{\pa_b u^{(k)} + q^{(k)}\cA^{(k)}_b}{T^{(k)}} \; ,
\nonu \\
{}
\lab{j-brane-A}
\er
where for each $k$-th \textsl{LL-brane}:
\br
{\bar g}^{(k)}_{ab} \equiv g^{(k)}_{ab} - \frac{1}{T_{(k)}^2}
(\pa_a u^{(k)} + q^{(k)}\cA^{(k)}_a)(\pa_b u^{(k)}  + q^{(k)}\cA^{(k)}_b) 
\nonu \\
g^{(k)}_{ab} = \pa_a X_{(k)}^\m G_{\m\n} \pa_b X_{(k)}^\n \quad , \quad 
\cA^{(k)}_a \equiv \pa_a X_{(k)}^\m A_\m \; .
\lab{ind-metric-ext-A-k}
\er
The \textsl{LL-brane} equations of motion have already been written down in
\rf{X-eqs-NG-A}--\rf{T-eq-NG-A} above.

Constructing wormhole solutions of static ``spherically-symmetric''-type 
\rf{static-spherical-EF} for the coupled gravity-gauge-field-\textsl{LL-brane} 
system \rf{gravity+GG+LL} proceeds through the following steps:

(i) Choose ``vacuum'' static ``spherically-symmetric''-type solutions
\rf{static-spherical-EF} of \rf{einstein+LL-eqs}--\rf{GG+LL-eqs}, \textsl{i.e.}, 
without the delta-function terms due to the \textsl{LL-branes}, in each 
space-time region (separate ``universe'') given by 
$\bigl(-\infty\! <\!\eta\!<\!\eta^{(1)}_0\bigr)\, ,\,
\bigl(\eta^{(1)}_0 \!<\!\eta \!<\! \eta^{(2)}_0\bigr) ,
\ldots, \bigl(\eta^{(N)}_0 \!<\!\eta \!<\!\infty\bigr)$
with common horizon(s) at $\eta=\eta^{(k)}_0$ ($k=1,\ldots ,N$).

(ii) Each $k$-th \textsl{LL-brane} automatically locates itself on the
horizon at $\eta=\eta^{(k)}_0$ according to ``horizon straddling'' property 
\rf{straddling} of \textsl{LL-brane} dynamics. It thus will play the role of a
wormhole ``throat'' between two neighboring ``universes''.

(iii) Match the discontinuities of the derivatives of the metric and
the gauge field strength \rf{static-spherical-EF} across each horizon at
$\eta=\eta^{(k)}_0$ using the explicit expressions for the \textsl{LL-brane} 
stress-energy tensor and charge current density \rf{T-brane-A}--\rf{j-brane-A}.

Taking into account \rf{static-spherical-EF}--\rf{straddling}, we obtain from
\rf{T-brane-A} the following expression for the energy-momentum tensor of each 
$k$-th \textsl{LL-brane} (here we suppress the index $(k)$):
\be
T_{(brane)}^{\m\n} = S^{\m\n}\,\d (\eta-\eta_0)
\lab{T-S-0}
\ee
with surface energy-momentum tensor:
\be
S^{\m\n} \equiv \frac{T}{b_0^{1/2}}\,
\( \pa_\t X^\m \pa_\t X^\n - 
b_0 G^{ij} \pa_i X^\m \pa_j X^\n 
\)_{v=\t,\,\eta=\eta_0,\,\th^i =\s^i} \; ,
\lab{T-S-brane} 
\ee
where $G_{ij} = C(\eta) h_{ij}(\th)$ (cf. \rf{static-spherical-EF}, here
$i,j = \th,\vp$). For the non-zero components of \rf{T-S-brane} (with lower indices)
and its trace we find:
\be
S_{\eta\eta} = \frac{T}{b_0^{1/2}} \quad ,\quad 
S_{ij} = - T b_0^{1/2} G_{ij} \quad ,\quad 
S^\l_\l = - 2Tb_0^{1/2} \; .
\lab{S-comp}
\ee
For the \textsl{LL-brane} charge current densities we get accordingly:
\be
j_{(\mathrm{brane-k})}^\m =
\d^\m_0 q^{(k)} \sqrt{\det\Vert G_{ij}\Vert}\,\d (\eta-\eta^{(k)}_0) \; .
\lab{j-0}
\ee

With the help of \rf{T-S-0}--\rf{j-0} and using again 
\rf{static-spherical-EF}--\rf{straddling} the matching relations for the
discontinuities at each horizon $\eta =\eta^{(k)}_0$ become
(cf. Refs.~\refcite{BR-WH-1,BR-WH-2}):

(A) Matching relations from Einstein eqs.\rf{einstein+LL-eqs}:
\be
\llb \pa_\eta A \rrb_{\eta^{(k)}_0} = - 16\pi T^{(k)} \sqrt{b^{(k)}_0} 
\quad,\quad 
\llb \pa_\eta \ln C \rrb_{\eta^{(k)}_0} = 
- \frac{8\pi}{\sqrt{b^{(k)}_0}} T^{(k)}
\lab{eqsys-1-2}
\ee
using notation $\bigl\lb Y \bigr\rb_{\eta_0} \equiv 
Y\bv_{\eta \to \eta_0 +0} - Y\bv_{\eta \to \eta_0 -0}$ for any quantity $Y$.

(B) Matching relations from nonlinear gauge field eqs.\rf{GG+LL-eqs}:
\be
\llb F_{v\eta} \rrb_{\eta^{(k)}_0} = q^{(k)} \; .
\lab{eqsys-4}
\ee

(C) The only non-trivial contribution of second-order \textsl{LL-brane} 
equations of motion \rf{X-eqs-NG-A} in the case of \textsl{LL-brane} coordinate
embedding \rf{X-embed} comes from the $X^0$-equation of motion which yields:
\be
\pa_0 T^{(k)} + 
\frac{T^{(k)}}{2} \( \llangle \pa_\eta A \rrangle_{\eta^{(k)}_0} 
+ 2 b^{(k)}_0 \llangle \pa_\eta \ln C \rrangle_{\eta^{(k)}_0} \)
-  \sqrt{b^{(k)}_0} q \llangle F_{v\eta}\rrangle_{\eta^{(k)}_0} = 0
\lab{eqsys-3}
\ee
with notation $\llangle Y \rrangle_{\eta_0} \equiv 
\h \Bigl( Y\bv_{\eta \to \eta_0 +0} + Y\bv_{\eta \to \eta_0 -0}\Bigr)$. 
In what follows we will take time-independent dynamical \textsl{LL-brane}
tension(s) ($\pa_0 T^{(k)}=0$) because of matching static bulk space-time 
geometries. Let us also note that the appearance of mean values of the
corresponding quantities with discontinuities across the horizons follows
the resolution of the discontinuity problem given in 
Ref.~\refcite{Israel-66} (see also Ref.~\refcite{BGG}).

The wormhole solutions presented in the next Section share the
following important properties:

(a) The \textsl{LL-branes} at the wormhole ``throats'' represent ``exotic'' 
matter with $T\leq 0$, \textsl{i.e.}, negative or zero brane tension implying 
violation of null-energy conditions as predicted by general wormhole arguments
\ct{visser-book+} (although the latter could be remedied via quantum fluctuations).

(b) The wormhole space-times constructed via \textsl{LL-branes} at
their ``throats'' are {\em not} traversable w.r.t. the ``laboratory'' time of a 
static observer in either of the different ``universes'' comprising the pertinent 
wormhole space-time manifold since the \textsl{LL-branes} sitting at the
``throats'' look as black hole horizons to the static observer. On the other hand, 
these wormholes {\em are traversable} w.r.t. the {\em proper time} of a 
traveling observer.

Proper-time traversability can be easily seen by considering dynamics of 
test particle of mass $m_0$ (``traveling observer'') in a wormhole background, 
which is described by the reparametrization-invariant world-line action:
\be
S_{\mathrm{particle}} = \h \int d\l \Bigl\lb\frac{1}{e}\xdot^\m \xdot^\n G_{\m\n}
- e m_0^2 \rb \; .
\lab{test-particle}
\ee
Using energy $\cE$ and orbital momentum $\cJ$ conservation and introducing the 
{\em proper} world-line time $s$ ($\frac{ds}{d\l}= e m_0$), the ``mass-shell''
constraint equation (the equation w.r.t. the ``einbein'' $e$) produced by the action 
\rf{test-particle}) yields:
\be
\(\frac{d\eta}{ds}\)^2 + \cV_{\mathrm{eff}} (\eta) = \frac{\cE^2}{m_0^2}
\quad ,\quad 
\cV_{\mathrm{eff}} (\eta) \equiv A(\eta) \Bigl( 1 + \frac{\cJ^2}{m_0^2 C(\eta)}\Bigr) 
\lab{particle-eq-2}
\ee
where the metric coefficients $A (\eta),\, C(\eta)$ are those in
\rf{static-spherical-EF}.
Irrespectively of the specific form of the ``effective potential'' in
\rf{particle-eq-2}, a ``radially'' moving (with zero ``impact'' parameter
$\cJ=0$) traveling observer (and with sufficiently large energy $\cE$) will
always cross within finite amount of proper-time through any ``throat'' 
($\eta = \eta_0^{(k)}$, where $A(\eta_0^{(k)})=0$) from one ``universe'' to another.

\section{Charge-Hiding Wormholes}
First we will construct ``one-throat'' wormhole solutions to \rf{gravity+GG+LL} with
the charged \textsl{LL-brane} occupying the wormhole ``throat'', which connects
a non-compact ``universe'' with Reissner-Nordstr{\"o}m-(anti)-de-Sitter-type geometry 
\rf{cornell-sol}--\rf{RN-dS+const-electr} (where the cosmological constant is 
partially or entirely {\em dynamically} generated) to a compactified 
(``tube-like'') ``universe'' of (generalized) Levi-Civita-Bertotti-Robinson type 
\rf{gen-BR-metric}--\rf{electr-static}. These wormholes possess
the novel property of {\em hiding} electric charge from external observer in the
non-compact ``universe'', \textsl{i.e.}, the whole electric flux produced by
the charged \textsl{LL-brane} at the wormhole ``throat'' is pushed into the 
``tube-like'' ``universe''. As a result, the non-compact ``universe'' becomes
electrically neutral with Schwarzschild-(anti-)de-Sitter or purely Schwarzschild
geometry.

We find several types of such wormhole solutions. The first one exists when 
the bare cosmological constant $\L > - 2\pi f^2$, in particular, when $\L$ is
absent from the very beginning, the whole effective cosmological constant 
being dynamically generated \rf{CC-eff}. This wormhole solution is
constructed as follows:

(A-1) ``Left universe'' of Levi-Civita-Bertotti-Robinson (``tube-like'') type 
with geometry $AdS_2 \times S^2$ \rf{AdS2} for $\eta< 0$:
\br
A(\eta) = 4\pi \( c_F^2 - \sqrt{2}f|c_F| - \frac{\L}{4\pi}\)\,\eta^2
\quad ,\quad C(\eta) \equiv r_0^2 = \frac{1}{4\pi c_F^2 + \L} = \mathrm{const} \; ,
\nonu\\
|F_{v\eta}| = |c_F| > \frac{f}{\sqrt{2}}\Bigl( 1+\sqrt{1+\frac{\L}{2\pi f^2}}\,\Bigr)
= \mathrm{const} \; ;
\lab{BR-AdS2-left-a}
\er

(A-2) Non-compact ``right universe'' for $\eta> 0$ comprising the exterior region of 
Reissner-Nordstr{\"o}m-de-Sitter-type black hole beyond the middle
(Schwarzschild-type) horizon $r_0$ (\textsl{cf.} \rf{cornell-sol}--\rf{CC-eff}):
\br
A(\eta) \equiv A_{\mathrm{RNdS}} (r_0 + \eta) = 
1 - \sqrt{8\pi}|Q|f - \frac{2m}{r_0 + \eta} + \frac{Q^2}{(r_0 + \eta)^2} 
- \frac{\L + 2\pi f^2}{3} (r_0 + \eta)^2 \; ,
\nonu \\
C(\eta) = (r_0 + \eta)^2 \quad ,\quad
F_{v\eta} = F_{0r} =
\frac{\vareps_F f}{\sqrt{2}} + \frac{Q}{\sqrt{4\pi}\, (r_0 + \eta)^2} \; .
\phantom{aaaaaaaa}
\lab{RNdS-right-a}
\er
Here $A(0) = A_{\mathrm{RNdS}} (r_0) = 0$ and
$\pa_\eta A(0) = \pa_r A_{\mathrm{RNdS}} (r_0) > 0$.

The next wormhole solution, which exists for large negative bare
cosmological constant $\L <0 \;,\; |\L| > 2\pi f^2$, is built by:

(B-1) The same type of ``left universe'' of Levi-Civita-Bertotti-Robinson
type with geometry $AdS_2 \times S^2$ \rf{AdS2} for $\eta< 0$ as in
\rf{BR-AdS2-left-a}:
\br
A(\eta) = 4\pi \( c_F^2 - \sqrt{2}f|c_F| + \frac{|\L|}{4\pi}\)\,\eta^2
\;\; ,\;\; C(\eta)  \equiv r_0^2 
= \frac{1}{4\pi c_F^2 - |\L|} = \mathrm{const} \; ,
\nonu\\
|F_{v\eta}| = |c_F| > \sqrt{\frac{1}{4\pi}|\L|} \; . \phantom{aaaaaaaa}
\lab{BR-AdS2-left-b}
\er

(B-2) Non-compact ``right universe'' for $\eta> 0$ comprising the exterior region of 
Reissner-Nordstr{\"o}m-{\em anti}-de-Sitter-type black hole beyond the outer 
(Schwarzschild-type) horizon $r_0$:
\br
A(\eta) \equiv A_{\mathrm{RN-AdS}} (r_0 + \eta) = 
1 - \sqrt{8\pi}|Q|f - \frac{2m}{r_0 + \eta} + \frac{Q^2}{(r_0 + \eta)^2} 
+ \frac{|\L| - 2\pi f^2}{3} (r_0 + \eta)^2 \; ,
\nonu \\
C(\eta) = (r_0 + \eta)^2 \quad ,\quad
F_{v\eta} = F_{0r} =
\frac{\vareps_F f}{\sqrt{2}} + \frac{Q}{\sqrt{4\pi}\, (r_0 + \eta)^2}
\; . \phantom{aaaaaaaa}
\lab{RN-AdS-right-b}
\er
Here again $A(0) = A_{\mathrm{RN-AdS}} (r_0) = 0$ and
$\pa_\eta A(0) = \pa_r A_{\mathrm{RN-AdS}} (r_0) > 0$.

For the special negative value of the bare cosmological constant $\L = - 2\pi f^2$
we find a third wormhole solution consisting of:

(C-1) The same type of ``left universe'' of Levi-Civita-Bertotti-Robinson
type with geometry $AdS_2 \times S^2$ \rf{AdS2} for $\eta< 0$ as in
\rf{BR-AdS2-left-a}:
\br
A(\eta) = 4\pi \( |c_F| - \frac{f}{\sqrt{2}}\)^2\,\eta^2
\quad ,\quad C(\eta)  \equiv r_0^2 = \frac{1}{4\pi \bigl(c_F^2 -\h f^2\bigr)} 
= \mathrm{const} \; ,
\nonu\\
|F_{v\eta}| = |c_F| > \frac{f}{\sqrt{2}} \; . \phantom{aaaaaaaa}
\lab{BR-AdS2-left-c}
\er

(C-2) Non-compact ``right universe'' for $\eta> 0$ comprising the exterior region of 
Reissner-Nordstr{\"o}m-type black hole beyond the outer (Schwarzschild-type) horizon $r_0$:
\br
A(\eta) \equiv A_{\mathrm{RN}} (r_0 + \eta) = 
1 - \sqrt{8\pi}|Q|f - \frac{2m}{r_0 + \eta} + \frac{Q^2}{(r_0 + \eta)^2} \; ,
\nonu \\
C(\eta) = (r_0 + \eta)^2 \quad ,\quad
F_{v\eta} = F_{0r} =
\frac{\vareps_F f}{\sqrt{2}} + \frac{Q}{\sqrt{4\pi}\, (r_0 + \eta)^2} \; .
\lab{RN-right-c}
\er
Here again $A(0) = A_{\mathrm{RN}} (r_0) = 0$ and
$\pa_\eta A(0) = \pa_r A_{\mathrm{RN}} (r_0) > 0$.

Substituting \rf{BR-AdS2-left-a}--\rf{RNdS-right-a}, 
\rf{BR-AdS2-left-b}--\rf{RN-AdS-right-b} and \rf{BR-AdS2-left-c}--\rf{RN-right-c}
into the set of matching relations \rf{eqsys-1-2}--\rf{eqsys-3} determines all 
parameters of the wormhole solutions
$\( r_0, m, Q, b_0, T\)$ in terms of $q$ (the \textsl{LL-brane} charge) and $f$
(coupling constant of the ``square-root'' Maxwell term in \rf{gravity+GG+LL}):
\br
Q=0 \;\; ,\;\; |c_F| = |q| + \frac{f}{\sqrt{2}} \;\;,\;\; 
\mathrm{sign}(q) = - \mathrm{sign}(F_{v\eta}) \equiv - \mathrm{sign}(c_F) \; ,
\lab{param-1}\\
\frac{1}{r_0^2} = 4\pi \Bigl(|q| + \frac{f}{\sqrt{2}}\Bigr)^2 + \L \quad ,\quad 
m = \frac{r_0}{2} \Bigl\lb 1 - \frac{1}{3}\bigl(\L + 2\pi f^2\bigr) r_0^2\Bigr\rb \; ,
\lab{param-2}\\
b_0 = \frac{1}{4}\( q^2 + \sqrt{2}f|q|\) 
\Bigl\lb \Bigl(|q|+\frac{f}{\sqrt{2}}\Bigr)^2 + \frac{1}{4\pi}\L \Bigr\rb^{-1}
\;\; ,\;\; T^2= \frac{1}{16\pi} \( q^2 + \sqrt{2}f|q|\) \; ,
\lab{param-3}
\er
and the bare cosmological constant must be in the interval:
\be
-4\pi\Bigl(|q|+\frac{f}{\sqrt{2}}\Bigr)^2 < \L < 4\pi\Bigl(q^2 -\frac{f^2}{2}\Bigr)
\; ,
\lab{CC-interval}
\ee
in particular, $\L$ could be zero.

The next wormhole solution has $Rind_2 \times S^2$ as compactified
``left'' universe whenever $\L > - 2\pi f^2$. It is built by:

(D-1) ``Left universe'' for $\eta <0$ of Levi-Civita-Bertotti-Robinson 
(``tube-like'') type with geometry $Rind_2 \times S^2$ \rf{Rindler2}:
\br
A(\eta) = -\eta \quad ,\quad 
C(\eta) \equiv r_0^2 = \frac{1}{4\pi c_F^2 + \L} = \mathrm{const} \; ,
\nonu\\
|F_{v\eta}| = |c_F| = 
\frac{f}{\sqrt{2}}\Bigl( 1 +\sqrt{1+\frac{\L}{2\pi f^2}}\;\Bigr) 
= \mathrm{const} \; ;
\lab{BR-AdS2-left-d}
\er

(D-2) Non-compact ``right universe'' for $\eta> 0$ comprising the exterior region of 
Reissner-Nordstr{\"o}m-de-Sitter-type black hole beyond the middle (Schwarzschild-type) 
horizon $r_0$ as in \rf{RNdS-right-a}.

Again substituting \rf{BR-AdS2-left-d} and \rf{RNdS-right-a} into the set of 
matching relations \rf{eqsys-1-2}--\rf{eqsys-3} determines all parameters of 
the wormhole solution (D-1)--(D-2) in complete analogy with
\rf{param-1}--\rf{param-3}:
\br
Q=0 \;\; ,\;\; |c_F| = |q| + \frac{f}{\sqrt{2}} \;\;,\;\; 
\mathrm{sign}(q) = - \mathrm{sign}(F_{v\eta}) \equiv - \mathrm{sign}(c_F) \; ,
\lab{param-1-R}\\
\L = 4\pi \Bigl( q^2 - \frac{f^2}{2}) \;\; ,\;\;
\frac{1}{r_0^2} = 8\pi \Bigl(q^2 + \frac{f}{\sqrt{2}}|q|\Bigr) \;\; ,\;\; 
m = \frac{r_0}{2} \Bigl\lb 1 - \frac{4\pi q^2}{3} r_0^2\Bigr\rb \; ,
\lab{param-2-R}\\
b_0 = \frac{1}{4}\Bigl\lb 1 + r_0 - 4\pi q^2 r_0^2\Bigr\rb 
\;\; ,\;\; T^2=\frac{b_0}{2\pi} \Bigl( q^2+\frac{f}{\sqrt{2}}|q|\Bigr) \; .
\lab{param-3-R}
\er

The result $Q=0$ in \rf{param-1} and \rf{param-1-R} has profound consequences. 
Namely, the absence of Coulomb field in spite of the presence of the charged
\textsl{LL-brane} source leads us to the following important observations:

(A) The ``right-universe'' in the wormhole solutions (A-1)--(A-2) 
(Eqs.\rf{BR-AdS2-left-a}--\rf{RNdS-right-a}) and (D-1)--(D-2) 
(Eqs.\rf{BR-AdS2-left-d}, \rf{RNdS-right-a}) becomes exterior region of 
electrically neutral Schwarzschild-de-Sitter 
black hole beyond the internal (Schwarzschild-type) horizon carrying a vacuum 
constant radial electric field $|F_{v\eta}|=|F_{0r}|=\frac{f}{\sqrt{2}}$.

(B) The ``right-universe'' in the wormhole solution (B-1)--(B-2) 
(Eqs.\rf{BR-AdS2-left-b}--\rf{RN-AdS-right-b}) becomes exterior 
region of electrically neutral Schwarzschild-{\em anti}-de-Sitter black hole beyond the
sole (Schwarzschild-type) horizon carrying a vacuum 
constant radial electric field $|F_{v\eta}|=|F_{0r}|=\frac{f}{\sqrt{2}}$.

(C) The ``right-universe'' in the wormhole solution (C-1)--(C-2) 
(Eqs.\rf{BR-AdS2-left-c}--\rf{RN-right-c}) becomes exterior 
region of the ordinary electrically neutral Schwarzschild black hole beyond
the horizon carrying a vacuum constant radial electric field 
$|F_{v\eta}|=|F_{0r}|=\frac{f}{\sqrt{2}}$.

(D) According to \rf{param-1} and \rf{param-1-R} the whole flux of the
electric field $|F_{0\eta}|$ with $|F_{0\eta}|=|F_{v\eta}|=\frac{f}{\sqrt{2}}+|q|$ 
produced by the \textsl{LL-brane} 
charge $q$ flows only into the compactified ``left universe'' of 
Levi-Civita-Bertotti-Robinson type ($AdS_2 \times S^2$ \rf{AdS2} or
$Rind_2 \times S^2$ \rf{Rindler2}). Due to the absence of electric flux
in the non-compact ``right universe'', an outside observer there will therefore
detect the charged \textsl{LL-brane} as a neutral object.

A clearer explanation of above statements (A)-(D) can be given if we recall that 
the electric flux is defined in terms of the
electric displacement field $\vec{D}$, which in the present case is significantly
different from the electric field $\vec{E}$ due to the presence of the ``square-root''
Maxwell term in \rf{gravity+GG+LL}:
\be
\vec{D} = \Bigl( 1 - \frac{f}{\sqrt{2}|\vec{E}|}\Bigr)\,\vec{E} \; .
\lab{dielectric-def}
\ee
Indeed, in the absence of magnetic field the $0$-th component of the
nonlinear gauge field Eqs.\rf{GG+LL-eqs} can be written in terms of $\vec{D}$ 
\rf{dielectric-def} as:
\be
\vec{\pa}\, .\(\sqrt{-G}\,\vec{D}\) - \sqrt{-G}\, J^0 = 0 \; ,
\lab{GG+LL-eqs-0}
\ee
where $J^\m = \frac{1}{\sqrt{-G}} j^\m$ is the charge vector current, so that:
\be
\int_{\pa \S} d\vec{S}.\vec{D} = Q_{\mathrm{total}} = \int_{\S} dV\, J^0 \; .
\lab{flux-eq}
\ee
Here the factors $\sqrt{-G}$ go into the definition of the corresponding 
volume forms (integration measures) on the three-dimensional region $\S$ and
its boundary $\pa \S$. 

Thus, Eq.\rf{flux-eq} tells us that the electric flux from the charged \textsl{LL-brane}
flowing into the non-compact ``right universes'', where the constant radial 
vacuum electric field has magnitude $|\vec{E}|=\frac{f}{\sqrt{2}}$, is {\em zero} 
since $\vec{D}=0$ there according to \rf{dielectric-def}. On the other hand,
inside the compactified Levi-Civita-Bertotti-Robinson-type ``left universe'':
\be
\vec{D} = \frac{|q|}{\frac{f}{\sqrt{2}}+|q|}\,\vec{E} = - q \hat{\eta} \; ,
\lab{dielectric-LCBR}
\ee
where $\hat{\eta}$ denotes the unit vector along the ``radial-like'' $\eta$
coordinate (here we have used relations \rf{param-1} and \rf{param-1-R}). 
Therefore, the whole electric flux from the charged \textsl{LL-brane}
is expelled into the ``tube-like'' ``left universe''.

The geometry of the charge-``hiding'' wormholes is visualized in Figure 1.

\begin{figure}
\begin{center}
\includegraphics{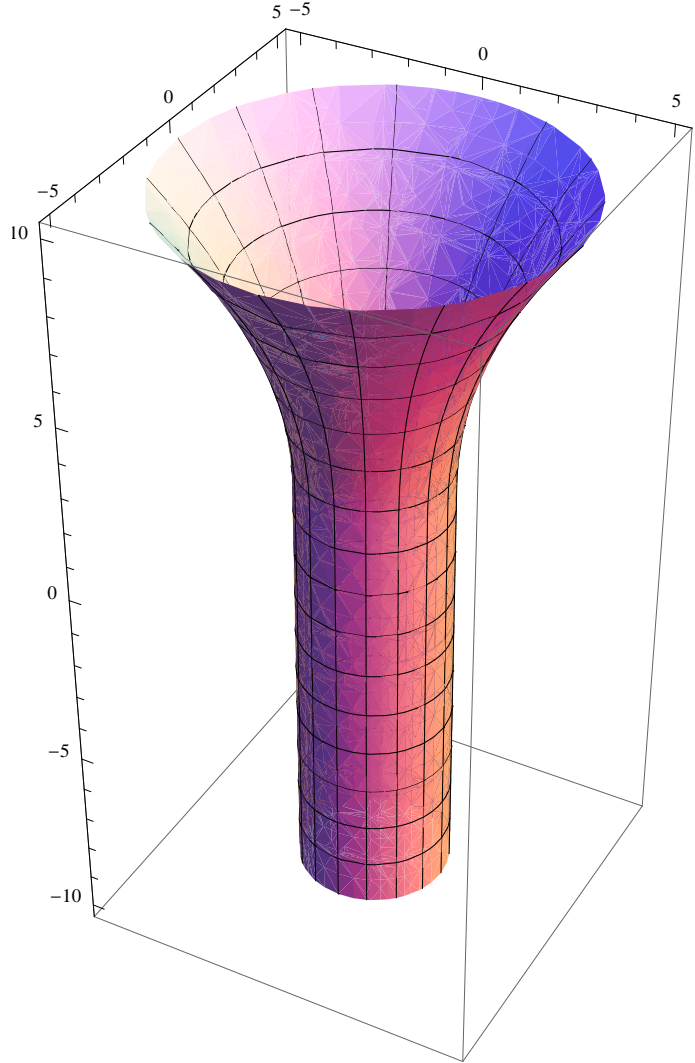}
\caption{Shape of $t=const$ and $\th=\frac{\pi}{2}$ slice of
charge-``hiding'' wormhole geometry. The whole electric flux is expelled into the
lower cylindric tube.}
\end{center}
\end{figure}

\section{Charge-Confining Wormhole}
Apart from the above {\em charge-hiding effect}  produced by ``one-throat'' wormhole
connecting a non-compact ``universe'' to a compactified ``tube-like''
``universe'' via \textsl{LL-brane} we find an even more interesting ``two-throat''
wormhole solution exhibiting {\em QCD-like charge confinement} effect. Namely, let us
now consider a self-consistent coupling of the gravity/nonlinear-gauge-field system 
\rf{gravity+GG} with two separate {\em oppositely charged}, but otherwise
identical \textsl{LL-branes} described by the action \rf{gravity+GG+LL}
and the resulting equations of motion \rf{einstein+LL-eqs}--\rf{ind-metric-ext-A-k} 
(here $N=2$, $T^{(1)}=T^{(2)}\equiv T$, $b_0^{(1)}=b_0^{(2)}\equiv b_0$,
$q^{(1)}\equiv q = - q^{(2)}$). 

Using the general scheme outlined in Section 5 we construct a solution where
the total ``two-throat'' wormhole space-time manifold is built as
follows:

(E-1) 
``Left-most'' non-compact ``universe'' comprising the exterior region of 
Reissner-Nordshtr{\"om}-de-Sitter-type black hole beyond the middle Schwarzschild-type 
horizon $r_0$ for the ``radial-like'' $\eta$-coordinate interval (see also
Eqs.\rf{middle-interval} and \rf{right-most-interval} below):
\be
-\infty < \eta < -\eta_0 
\equiv - \Bigl\lb 4\pi\(\sqrt{2}f|c_F| - c_F^2\) + \L \Bigr\rb^{-\h} \; ,
\lab{left-most-interval}
\ee
where (\textsl{cf.} \rf{cornell-sol}--\rf{CC-eff}):
\br
A(\eta) = A_{\mathrm{RNdS}}(r_0 - \eta_0 - \eta)
\nonu \\
= 1 - \sqrt{8\pi}|Q|f - \frac{2m}{r_0 - \eta_0 - \eta} + 
\frac{Q^2}{(r_0 - \eta_0 - \eta)^2} 
- \frac{\L + 2\pi f^2}{3} (r_0 - \eta_0 - \eta)^2 \; ,
\lab{RNdS-left-most-1} \\
C(\eta) = (r_0 - \eta_0 - \eta)^2 \quad ,\quad
F_{v\eta}(\eta) = F_{0r}(\eta) = \frac{\vareps_F f}{\sqrt{2}} 
+ \frac{Q}{\sqrt{4\pi}\,(r_0 - \eta_0 - \eta)^2} \; .
\lab{RNdS-left-most-2}
\er
Here $A (-\eta_0) = A_{\mathrm{RNdS}}(r_0) = 0$ and
$\pa_\eta A (-\eta_0) = - \pa_r A_{\mathrm{RNdS}}(r_0) < 0$.

(E-2) ``Middle'' compactified ``tube-like'' ``universe'' of
Levi-Civita-Bertotti-Robinson type with geometry $dS_2 \times S^2$ 
\rf{dS2}--\rf{dS2-cF-1} comprising the finite extent (w.r.t. $\eta$-coordinate) 
region between the two horizons of $dS_2$ at $\eta = \pm \eta_0$:
\be
-\eta_0 < \eta < \eta_0 
\equiv \Bigl\lb 4\pi\(\sqrt{2}f|c_F| - c_F^2\) + \L \Bigr\rb^{-\h} \; ,
\lab{middle-interval}
\ee
where (cf. Eqs.\rf{dS2}--\rf{dS2-cF-1}):
\br
A(\eta) = 1 - \Bigl\lb 4\pi\(\sqrt{2}f|c_F| - c_F^2\) + \L \Bigr\rb\,\eta^2
\quad ,\quad A(\pm \eta_0) = 0 \; ,
\lab{LCBR-middle-1} \\
C(\eta) = r_0^2 = \frac{1}{4\pi c_F^2 + \L} \quad ,\quad
|F_{v\eta}| = |c_F| < 
\frac{f}{\sqrt{2}}\Bigl( 1 + \sqrt{1 + \frac{\L}{2\pi f^2}}\Bigr) \; ,
\lab{LCBR-middle-2}
\er
with $\L > - 2\pi f^2$;

(E-3) 
``Right-most'' non-compact ``universe'' comprising the exterior region of 
Reissner-Nordshtr{\"om}-de-Sitter-type black hole beyond the middle Schwarzschild-type 
horizon $r_0$ for the ``radial-like'' $\eta$-coordinate interval:
\be
\eta_0 < \eta < \infty \quad (\eta_0 \; \mathrm{as ~in ~\rf{middle-interval}}) \; ,
\lab{right-most-interval}
\ee
\textsl{i.e.}, a mirror image ($-\eta \to \eta$) of the ``left-most'' ``universe''
\rf{RNdS-left-most-1}--\rf{RNdS-left-most-2}:
\br
A(\eta) = A_{\mathrm{RNdS}}(r_0 + \eta - \eta_0)
\nonu \\
= 1 - \sqrt{8\pi}|Q|f - \frac{2m}{r_0 + \eta - \eta_0} + 
\frac{Q^2}{(r_0 + \eta - \eta_0)^2} 
- \frac{\L + 2\pi f^2}{3} (r_0 + \eta - \eta_0)^2 \; ,
\lab{RNdS-right-most-1} \\
C(\eta) = (r_0 + \eta - \eta_0)^2 \quad ,\quad
F_{v\eta}(\eta) = F_{0r}(\eta) = \frac{\vareps_F f}{\sqrt{2}} 
+ \frac{Q}{\sqrt{4\pi}\,(r_0 + \eta - \eta_0)^2} \; .
\lab{RNdS-right-most-2}
\er
Here $A (\eta_0) = A_{\mathrm{RNdS}}(r_0) = 0$ and
$\pa_\eta A (\eta_0) = \pa_r A_{\mathrm{RNdS}}(r_0) > 0$.

According to the ``horizon straddling'' property \rf{straddling} of world-volume
\textsl{LL-brane} dynamics, 
each one of the two charged \textsl{LL-branes} (with equal world-volume parameters
$(T,b_0)$ but with opposite charges $\pm q$, cf. \rf{LL-action+EM}), 
automatically locates itself on one of the two common 
horizons between ``left-most'' (E-1) and middle (E-2) ``universes'' at 
$\eta = - \eta_0$ and between middle (E-2) and ``right-most'' (E-3) ``universes'' 
at $\eta = \eta_0$, respectively. 

Now, as we did in the previous Section, substituting 
\rf{left-most-interval}--\rf{RNdS-right-most-2} 
into the set of matching relations \rf{eqsys-1-2}--\rf{eqsys-3} determines all 
parameters of the wormhole solutions
$\( r_0, \eta_0, m, Q, b_0, T\)$ in terms of $|q|$ (the magnitude of the \textsl{LL-brane}
charges) and $f$ (the coupling constant of the ``square-root'' Maxwell term in 
\rf{gravity+GG+LL}):
\br
Q=0 \;\; ,\;\; |c_F| = |q| + \frac{f}{\sqrt{2}} \;\;,\;\; 
\mathrm{sign}(q) = - \mathrm{sign}(F_{v\eta}) \equiv - \mathrm{sign}(c_F) \; ,
\lab{param-1-conf}\\
\frac{1}{r_0^2} = 4\pi \Bigl(|q| + \frac{f}{\sqrt{2}}\Bigr)^2 + \L \quad ,\quad 
m = \frac{r_0}{2} \Bigl\lb 1 - \frac{1}{3}\bigl(\L + 2\pi f^2\bigr) r_0^2\Bigr\rb \; ,
\lab{param-2-conf}\\
\eta_0 = \Bigl\lb 2\pi (f^2 - 2q^2) + \L\Bigr\rb^{-\h} \; ,
\lab{param-3-conf} \\
b_0 = \frac{1}{4}\( 1 - \bigl(\L + 2\pi f^2\bigr) r_0^2 +
2r_0 \sqrt{\L + 2\pi f^2 - 4\pi q^2}\) \; ,
\nonu\\
T^2= \frac{b_0}{4\pi}
\Bigl\lb \Bigl(|q|+\frac{f}{\sqrt{2}}\Bigr)^2 + \frac{1}{4\pi}\L \Bigr\rb  \; .
\lab{param-4-conf}
\er
The bare cosmological constant must be in the interval:
\be
\L \leq 0 \quad ,\quad |\L| < 2\pi (f^2 - 2 q^2) 
\quad \to \quad |q| < \frac{f}{\sqrt{2}} \; ,
\lab{CC-interval-conf}
\ee
in particular, $\L$ could be zero.

Again, as in the previous Section, relations \rf{param-1-conf} are of
primary importance. They tell us that:
\begin{itemize}
\item
The ``left-most'' \rf{left-most-interval}--\rf{RNdS-left-most-2} and ``right-most''
\rf{right-most-interval}--\rf{RNdS-right-most-2} non-compact ``universes'' become
two identical copies of the {\em electrically neutral} exterior region of
Schwarzschild-de-Sitter black hole beyond the Schwarzschild horizon
carrying a constant vacuum radial electric field with magnitude 
$|F_{v\eta}|=|F_{0r}|=\frac{f}{\sqrt{2}}$ pointing inbound towards the
horizon in one of these ``universes'' and pointing outbound w.r.t. the horizon
in the second ``universe''. The corresponding electric displacement field 
$\vec{D}=0$, so there is {\em no} electric flux there 
(cf. Eq.\rf{dielectric-def}--\rf{flux-eq}).
\item
The whole electric flux produced by the two charged \textsl{LL-branes} with 
opposite charges $\pm q$ at the boundaries of the above non-compact ``universes''
is {\em confined} within the ``tube-like'' middle ``universe''
\rf{middle-interval}--\rf{LCBR-middle-2} where the constant electric field is
$|F_{v\eta}|=\frac{f}{\sqrt{2}} + |q|$ with associated non-zero electric displacement
field $|\vec{D}|= |q|$ (cf. Eqs.\rf{dielectric-def}--\rf{flux-eq}). 
\end{itemize}

The geometry of the charge-confining wormhole is visualized in the Figure 2.

\begin{figure}
\begin{center}
\includegraphics{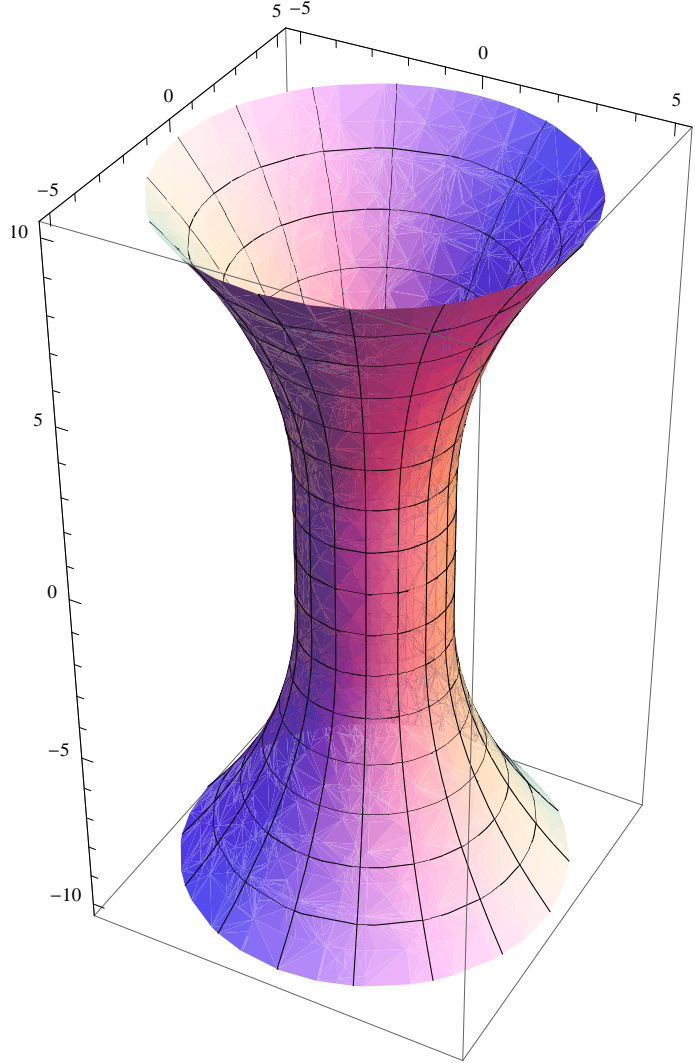}
\caption{Shape of $t=const$ and $\th=\frac{\pi}{2}$ slice of
charge-confining wormhole geometry. The whole electric flux is confined
within the middle cylindric tube.}
\end{center}
\end{figure}

\section{Discussion and Conclusions}
In this paper we have studied bulk gravity/nonlinear-gauge-field system
self-consistently coupled to one or two charged lightlike branes as matter
sources. An important feature of this system is the special form of the
nonlinear gauge field sector in \rf{gravity+GG+LL} previously known to produce QCD-like
confining dynamics in flat space-time \ct{GG-2}${}^{-}$\ct{GG-6}. 
The main objective here was to search for similar charge confining behavior in 
curved space-time, where the role of charged objects subject to confinement is 
played by charged lightlike branes. 

We found that charge-confining or charge-``hiding'' effects take place
within {\em wormhole} solutions to the coupled 
gravity/nonlinear-gauge-field/lightlike-brane system \rf{gravity+GG+LL} with
the following special structure:

(i) One of the ``universes'' comprising the total wormhole space-time
manifold must be a compactified ``universe'' of Levi-Civita-Bertotti-Robinson
(``tube-like'') type with geometry $\cM_2 \times S^2$ where the two-dimensional 
manifold $\cM_2$ possesses at least one horizon;

(ii) The one or two outer ``universe(s)'' are non-compact spherically symmetric
with at least one Schwarzschild-type horizon;

(iii) The matching (gluing together) of the compactified ``universe'' with
the (one of the two) outer non-compact ``universe(s)'' takes place at a
common horizon of both of them, which is automatically occupied by (one of the
participating) charged lightlike brane(s) (``horizon straddling'' as
dictated by world-volume lightlike brane dynamics);

(iv) Due to the presence of the ``square-root'' Maxwell term in \rf{gravity+GG+LL}
a non-zero constant vacuum electric field is generated in (any of) the outer
non-compact ``universe(s)'', however, the total flux is {\em zero} there
because of vanishing of the pertinent electric displacement field, so that
the charged lightlike brane occupying the ``throat'' between the non-compact
and the compactified ``tube-like'' ``universes'' appears as electrically
neutral to an external observer in the non-compact ``universe''.

(v) In the compactified ``tube-like'' ``universe'' the charged lightlike
brane(s) at the its ``border(s)'', where it is matched to the non-compact
``universe(s)'', produce a {\em non-zero} flux entirely confined within the
``tube-like'' ``universe'';

(vi) When only one charged lightlike brane is present, the compactified 
``tube-like'' ``universe'' with geometries $\cM_2 \times S^2$, where
$\cM_2=AdS_2$ or $\cM_2=Rind_2$, has an infinite extent w.r.t. ``radial-like''
$\eta$-coordinate of $\cM_2$ and it absorbs entirely the whole flux produced by 
the brane at the ``border''. In this way it hides the charge of the brane from an
outside observer in the ``neighboring'' non-compact ``universe''.

(vii) When two oppositely charged but otherwise identical lightlike branes
are present, the middle ``tube-like'' ``universe'' stretching between them  has geometry 
$dS_2 \times S^2$ and has final extent w.r.t. ``radial-like'' $\eta$-coordinate of the
$dS_2$ component. It absorbs entirely the whole flux produced between the branes at
its ``borders'', \textsl{i.e.}, the whole flux is {\em confined} within the
finite-extent ``tube-like'' region without flowing into any of the outside
non-compact space-time regions. 

It is natural to expect that in a confining theory the gauge field prefers
flux-tube configurations, however, the mathematical details of how this
happens might be complicated in flat space-time. On the other hand, in the
present curved space-time model we obtain the following clear and simple picture:

(a) Due to the presence of lightlike brane(s) as material source(s) of
gravity and gauge forces, the very special lightlike brane world-volume
dynamics {\em triggers} one or more transitions between non-compact and compactified
``tube-like'' space-time regions in the form of special wormhole
configurations with the lightlike brane(s) sitting at the ``throat(s)'';

(b) Again the special lightlike brane world-volume dynamics in combination with
the special properties of the additional ``square-root'' Maxwell term in the
nonlinear gauge field action cause the whole flux generated by the charged 
branes to be entirely confined within the compactified ``tube-like'' region.

As a final remark,
returning to the non-linear gauge field Eqs.\rf{GG-eqs} we see that
there exists a more general {\em vacuum} solution of the latter {\em without} the
assumption of staticity and spherical symmetry:
\be
F^2 \equiv F_{\k\l} F_{\m\n} G^{\k\m} G^{\l\n} = - f^2 = \mathrm{const} \; .
\lab{const-F2}
\ee
The latter automatically produces via Eq.\rf{stress-tensor-F} an effective positive
cosmological constant:
\be
T^{(F)}_{\m\n} = - \frac{f^2}{4} G_{\m\n} \quad ,\;\; \mathrm{i.e.} \;\;
\L_{\mathrm{eff}} = 2\pi f^2 \; .
\lab{stress-tensor-F-vac}
\ee
This reduces the gravity/gauge-field equations of motion \rf{einstein-eqs}, 
\rf{GG-eqs} to the vacuum Einstein equations (with effective cosmological
constant):
\be
R_{\m\n} - \h G_{\m\n} R + (\L + 2\pi f^2) G_{\m\n} = 0
\lab{einstein-eqs-vac}
\ee
supplemented with the constraint Eq.\rf{const-F2}.
Thus, 
assuming absence of magnetic field ($F_{mn}=0$), \textsl{i.e.}, 
$F^2 = 2 E_m E_n G^{mn} G^{00}\; ,\; E_m \equiv F_{0m}\; (m,n=1,2,3)$, we obtain the 
above described electrically neutral Schwarzschild-(anti)-de-Sitter or purely
Schwarzschild solutions with a constant vacuum electric field \rf{S-dS+const-electr}, 
which according to \rf{const-F2} has constant magnitude:
\be
|\vec{E}| \equiv \sqrt{-\h F^2} = \frac{f}{\sqrt{2}} \; ,
\lab{const-magnitude}
\ee
but it may point in arbitrary direction. In this vacuum with disordered 
constant-magnitude 
electric field it will not be able to pass energy to a test charged particle,
which instead will undergo a kind of Brownian motion, therefore {\em no} Schwinger
pair-creation mechanism will take place.

\section*{Acknowledgments}
E.N. and S.P. are supported by Bulgarian NSF grant \textsl{DO 02-257}.
Also, all of us acknowledge support of our collaboration through the exchange
agreement between the Ben-Gurion University 
and the Bulgarian Academy of Sciences.
We are grateful to Stoycho Yazadjiev for constructive discussions and Doug Singleton 
for correspondence. 
Thanks are also due to the referee for useful remarks.

\end{document}